# Laser cooling of positronium

K. Shu[1,2], Y. Tajima[2], R. Uozumi[2], N. Miyamoto[2], S. Shiraishi[2], T. Kobayashi[2], A. Ishida[3*], K. Yamada[3], R. W. Gladen[3], T. Namba[4], S. Asai[3], K. Wada[5], I. Mochizuki[5], T. Hyodo[5], K. Ito[6], K. Michishio[6], B. E. O'Rourke[6], N. Oshima[6], K. Yoshioka[1,2*]

[1]Photon Science Centre, School of Engineering, The University of Tokyo, 2-11-16 Yayoi, Bunkyo-ku, Tokyo 113-8656, Japan.
[2]Department of Applied Physics, School of Engineering, The University of Tokyo, 7-3-1 Hongo, Bunkyo-ku, Tokyo 113-8656, Japan.
[3]Department of Physics, Graduate School of Science, The University of Tokyo, 7-3-1 Hongo, Bunkyo-ku, Tokyo 113-0033, Japan.
[4]International Centre for Elementary Particle Physics (ICEPP), The University of Tokyo, 7-3-1 Hongo, Bunkyo-ku, Tokyo 113-0033, Japan.
[5]Institute of Materials Structure Science, High Energy Accelerator Research Organization (KEK), Tsukuba, Ibaraki 305-0801, Japan.
[6]National Institute of Advanced Industrial Science and Technology (AIST), Tsukuba-Central 2, 1-1-1 Umezono, Tsukuba, Ibaraki 305-8568, Japan.

**When laser radiation is skilfully applied, atoms and molecules can be cooled[1-3] allowing precise measurements and control of quantum systems. This is essential in fundamental studies of physics as well as practical applications such as precision spectroscopy[4-7], quantum-statistical-property manifesting ultracold gases[8-10], and quantum computing. In laser cooling, repeated cycles of laser photon absorption and direction-independent spontaneous emission can slow atoms and molecules to otherwise unattainable velocities. Simple systems can provide a rigorous testing ground for fundamental theories of physics; one such system is the purely leptonic positronium[11], an exotic atom of an electron and its antiparticle, the positron. However, the cooling of positronium has hitherto remained unrealised. Here, we demonstrate laser cooling of positronium. A novel laser system of a train of broadband pulses with successively increasing central frequencies was used to overcome major challenges presented by the short lifetime of positronium and the significant Doppler broadening and recoil as a consequence of its very light mass. One-dimensional chirp cooling of the dilute positronium gas in a counter-propagating configuration gave a final velocity distribution corresponding to approximately 1 K in a short time of 100 ns. This study on a pure leptonic system is**

**a major step in the field of low-temperature fundamental physics of antimatter, and is complementary to the laser cooling of antihydrogen[12], a hadron-containing exotic atom. Progress in this field is vital in elucidating the origin of the matter–antimatter asymmetry in the universe[13,14]. The application of laser cooling to positronium may afford a unique opportunity to rigorously test bound-state quantum electrodynamics. Moreover, laser cooling of positronium is key to the realisation of Bose–Einstein condensation in this matter–antimatter system.**

The impact of positronium (Ps) cooling on fundamental physics is significant. Ps is the simplest atom, a bound state of two leptons. It constitutes a rigorous testing ground for quantum electrodynamics[15], which is the most accurate theory in modern physics. An example is the 1S–2S transition frequency[16-18], where reduction of the measurement uncertainty is now required for a rigorous comparison with QED calculations[19,20]. Reduction of systematic errors using a cold Ps gas and an optical frequency comb is essential. Cold Ps atoms can provide a valuable experimental platform for searching for CPT symmetry breaking in the lepton sector, and investigating the effects of gravity on antimatter[13,14,21-26]. Currently, Ps is the only antiparticle-containing atomic system that can be densified to a satisfactory level in a gas phase. Owing to its light mass, only twice that of an electron, Bose–Einstein condensation (BEC) is expected to occur at relatively high temperatures[27-31] compared to ordinary atoms. Discussions regarding the generation of coherent γ-rays utilising Ps BECs have been noted[32,33]. The cooling of Ps is a requisite for such investigations.

Despite these high expectations, there has been no demonstration of laser cooling of Ps. The application of laser cooling of Ps using the 1S–2P transition with a natural linewidth of 50 MHz was investigated some 30 years ago[34]. There have been preliminary experimental implementation studies[35,36]. Cold Ps is also requisite for generating cold antihydrogen pulses[37], and laser cooling with a wide spectral width and long laser duration in a high magnetic field has recently been investigated theoretically for that purpose[38].

Laser cooling is an established technique applied to atoms and molecules. The elementary process of laser cooling is the reduction of the translational momentum of a moving particle, resulting from absorption of a laser photon propagating counter to the motion and the subsequent emission of a photon in a random direction. Cooling is achieved through the repeated cycles of absorption-emission. Depending on the experimental purpose and applicability, appropriate choices can be made from existing laser cooling methods such as the Doppler cooling, chirp cooling, magneto-optical traps, and Zeeman slowers.

Two major characteristics of Ps make the laser cooling challenging: its short lifetime (triplet 1S state of 142 ns) and the large recoil-induced frequency shift. The lifetime originates from the pair-annihilation into γ-rays, and the decay becomes faster in a magnetic field. Because Ps is nearly three orders of magnitude lighter than a hydrogen atom, the velocity change associated with the absorption and emission of photons is concomitantly large. The recoil velocity $v_r$ due to a photon of a wavelength of 243 nm, which induces the 1S–2P transition, is approximately $v_r = 1.5 \times 10^3$ m/s. The corresponding change in the resonance frequency due to the Doppler effect is 6.2 GHz, which is significantly larger than the natural linewidth. Therefore, laser cooling halts after one cooling cycle if we use a single-frequency cooling laser that is commonly used in the Doppler cooling of ordinary atoms. In addition, the recoil limit temperature is higher than the Doppler limit temperature, which is in contrast to ordinary atoms and molecules. The Doppler broadening of the 1S–2P transition at 300 K extends to approximately 460 GHz at the full width at half maximum (FWHM). In principle, cooling of Ps is possible by using a laser pulse with a linewidth of the order of 100 GHz for a long duration, but the resulting temperature is limited to a few 10 K by the linewidth.

Here, we extend the chirp cooling technique[39,40] that is used to decelerate atomic beams. In chirp cooling, the frequency of the laser light changes over time to follow the change in the shifted resonance frequency due to deceleration, maintaining the cooling cycle. If the laser used has a very fast frequency chirp, which is adapted to the large recoil-induced shift, significant Doppler broadening and short lifetime, Ps can be cooled down to the recoil-limited velocity distribution equivalent to sub-kelvin levels[41].

We used a previously demonstrated light source[41,42] that potentially slows the velocity distribution of Ps to near the recoil limit temperature by applying the chirp cooling technique. Using this tailored laser (see Fig. 1a), short optical pulses were successively output every 4.2 ns. Each pulse was spectrally continuous and had a bandwidth of 8.9 GHz at FWHM, simultaneously covering all 1S–2P transition frequencies split over 9.87 GHz. Thus, the pulse serves as both a cooling and repump laser pulse. The repump prevents the reduction in cooling efficiency owing to the spin polarisation of the 1S state during cooling. The central frequency of the pulse varies with each pulse, and the chirp is linear with a rate of $4.9 \times 10^2$ GHz/μs. We denote here this chirped train of optical pulses as a cooling laser. By performing chirp cooling in a counter-propagating configuration, we expect a velocity distribution close to the recoil-limited one after the progressive deceleration from fast to slow Ps via the chirp cooling mechanism. This laser cooling method realises slow Ps atoms in both electrostatic- and magnetostatic-field-free environments, which is an important step forward in precision spectroscopy.

The experimental configuration inside the vacuum chamber is shown in Fig. 1b and 1c. Here, we depict the pulsed generation of Ps gas in vacuum, three spatially superimposed laser beams interacting with Ps, and the acquisition of excitation signals for the 1S–2P transitions. The experiment was conducted at the Slow Positron Facility (SPF) of the Institute of Materials Structure Science (IMSS), the High Energy Accelerator Research Organization (KEK), Japan. Bunches of positrons[43] were injected at room temperature into a silica aerogel, which was used as a Ps formation medium[44]. Approximately $3 \times 10^3$ thermalised Ps atoms per positron bunch were released from the silica aerogel into the vacuum. Throughout this study, we evaluated the velocity distribution at 125 ns after production. At this point, we estimated that the Ps gas had a spatial spread of nearly 10 mm in the longitudinal and transverse directions, resulting in a density of approximately $10^3$ cm$^{-3}$. We introduced three collimated laser beams into the vacuum chamber, expanded to encompass the entirety of the spatial spread of Ps.

The three laser beams consisted of a cooling laser pulse train at 243 nm, a nanosecond laser pulse at 243 nm for velocity distribution measurements via the 1S–2P transition, and a nanosecond laser pulse at 532 nm to ionise the 2P state. We evaluated the velocity distribution of the Ps by Doppler spectroscopy using the 1S–2P transition. Herein, we refer to a laser pulse with a wavelength of 243 nm, which induces a velocity-sensitive 1S–2P transition, as the probe pulse. The Ps atoms prepared in the 2P state by the probe pulse were photoionised by a 532 nm nanosecond pulse before relaxation to the 1S state. The photoionised positrons produced were collected by a microchannel plate (MCP) that was driven in a pulsed manner after the end of all laser pulses, and the output current was converted to a voltage signal. This positron detection method offers high efficiency, allowing measurements within a limited experimental beam time. In addition, the environment was electrostatic-field-free until the interaction between Ps and the laser pulses was complete. The maximum magnitude of the residual static magnetic flux density was 0.15 mT over the entire experimental setup; thus, its influence on the lifetime of Ps was negligible.

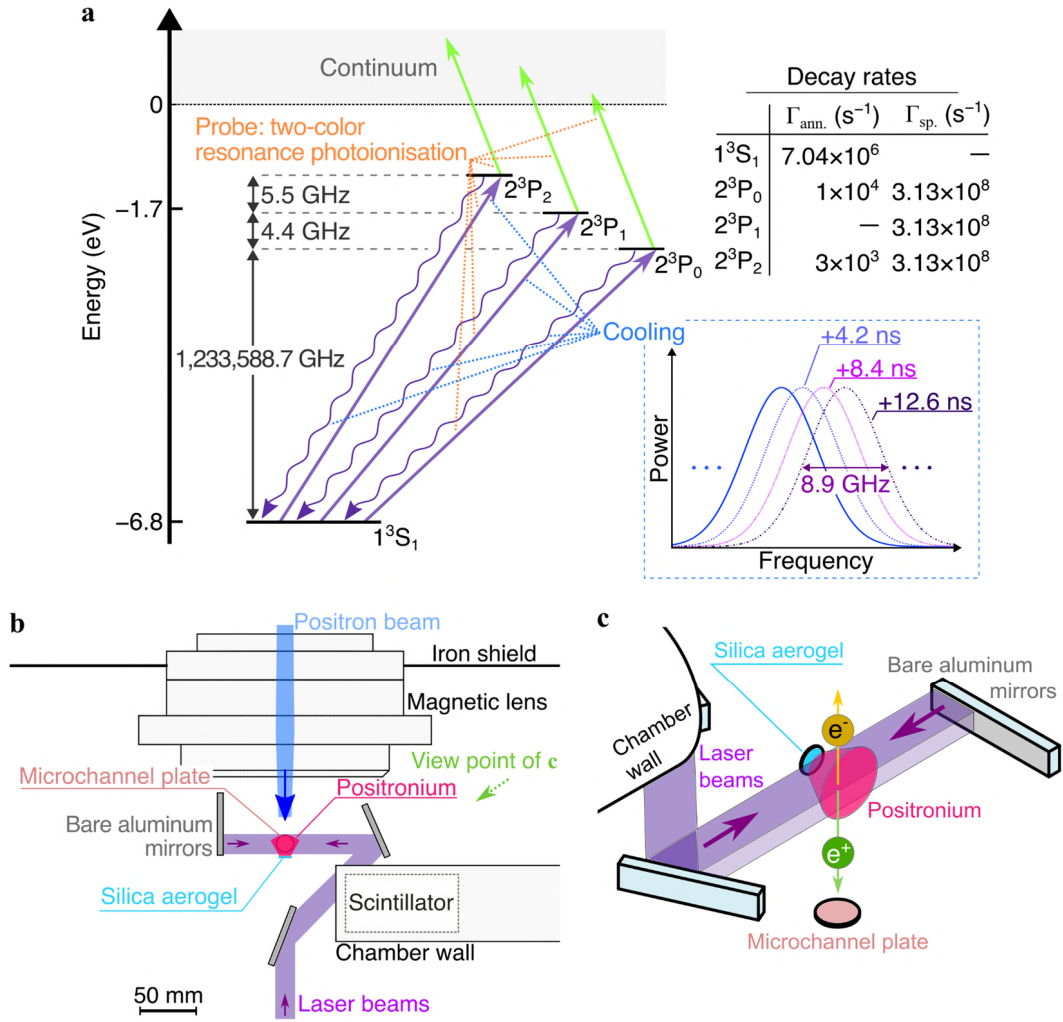

**Fig. 1: Experimental setup. a**, Interactions between the laser pulses and Ps. Relevant energy levels are shown up to the total angular momentum. The cooling pulses address all the 2P sublevels simultaneously, followed by the relaxation to the 1S level by spontaneous emission. In the Doppler profile measurements, we excite all the 2P sublevels simultaneously by the probe pulse. Right after the excitation, an ionisation laser pulse at 532 nm produces photoionised electrons and positrons from the excited Ps atoms in the 2P state. The frequencies shown in the figure correspond to the difference in eigen energies[45]. The table shows the decay rate for each state. $\Gamma_{\text{ann.}}$ and $\Gamma_{\text{sp.}}$ are the decay rates for the annihilation and the spontaneous emission, respectively. The figure enclosed in the dashed line depicts how the spectrum of the constituent pulses of the cooling laser changes with time. **b**, Top view of the experimental setup in the vacuum chamber. **c,** Bird's eye view of the setup.

Before proceeding to the laser cooling experiment, we examined the Ps gas temperature at 125 ns after the production of Ps. To evaluate temperature, we measured the Doppler profile in the 1S–2P excitation spectrum that reflects the velocity distribution due to the Doppler effect. The Doppler profile was obtained by evaluating the excitation signal originating from the ionised positrons while varying the optical frequency of the pulsed laser that induced the 1S–2P transition (see Methods). The measurement time for each wavelength was approximately 20 minutes. The measured Doppler profile is shown in Fig. 2a. It has a spread around the theoretically known frequency difference between the $1^3S_1$ and $2^3P_2$ levels (1,233,599 GHz), reflecting the velocity distribution of Ps atoms in the 1S state. The gas temperature was evaluated from the observed distribution by assuming that the transverse one-dimensional velocity distribution of Ps emitted from the generating source follows the Maxwell–Boltzmann distribution function. Fitting the data with a model that accounts for the spectral width of the 1S–2P transition-inducing laser pulse and Lamb dip associated with the counter-propagating laser beam revealed the gas temperature as $6.0 \times 10^2$ K. The Ps atoms in the aerogel were released into vacuum before thermalisation was complete, which was also observed in our time-of-flight measurements with the same Ps formation material and accounts for the temperature difference between the Ps gas and aerogel.

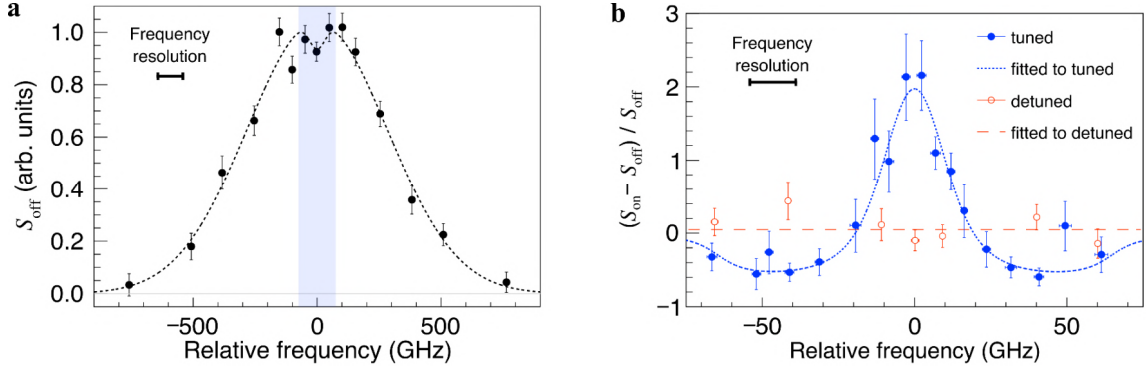

**Fig. 2: Demonstration of laser cooling. a**, Doppler profile of Ps in vacuum. The horizontal axis is the relative optical frequency of the laser pulse inducing the 1S–2P transition, which is evaluated with respect to the frequency difference between the $1^3S_1$ and $2^3P_2$ levels of 1,233,599 GHz. The vertical axis shows the 1S–2P excitation signal ($S_{off}$) without the cooling laser. Error bars originate from statistical and systematic errors in the measurement. The dotted curve is a fitting based on the Maxwell–Boltzmann distribution. We took into account the spectral width of the laser and the Lamb dip in this fitting function. The shaded area indicates the range of optical frequencies where we performed laser cooling and detected changes in the Doppler profile. **b**, Fractional change in velocity distribution of Ps as a result of laser cooling. We evaluated the fractional change by comparing the 1S–2P excitation signals with ($S_{on}$) and without ($S_{off}$) irradiation of the cooling laser. Filled circles represent the results when the optical frequencies of the cooling laser were set to a detuning suitable for cooling. The open circles represent the results of a control experiment with a larger detuning of the cooling laser. The dotted curve is a typical theoretical fitting to the data based on a phenomenological model. The spectral width of the probe pulse used in the fitting was 16 GHz. The dashed line is a constant function fitted to the data in a control experiment.

To demonstrate laser cooling, we irradiated the cooling laser for 100 ns, starting with the peak time of the positron pulse injection into the aerogel. The range of the Doppler shift of Ps affected by the cooling laser was approximately ±60 GHz with respect to the central frequency. Here, we measured the Doppler profile with an improved frequency resolution in the region described above (shaded area in Fig. 2a). We obtained an optical frequency resolution ranging from 8 to 16 GHz by narrowing the spectrum of the broadband laser pulse of a linewidth of 110 GHz (see Methods).

We measured the change in the Ps velocity distribution resulting from irradiation with the cooling laser. Figure 2b shows the fractional changes in the Doppler distribution. The excitation signal of the 1S–2P transition after irradiation with the cooling laser is defined as $S_{on}$, and the fractional change as $(S_{on} - S_{off})/S_{off}$. The filled circles show the results obtained when we tuned the optical frequency of the cooling laser to the conditions expected to produce a high cooling efficiency. The optical frequency (detuning) was adjusted to chirp from approximately 1,233,540 GHz (−59 GHz) to 1,233,590 GHz (−9 GHz) during the 100 ns duration of the cooling laser. The data showed a decrease in the number of Ps in the optical frequency range swept by the cooling laser, accompanied by an approximately threefold increase in the number of Ps near zero velocity within a narrow velocity range, demonstrating the laser cooling of Ps — this result is the anticipated breakthrough achieved using our laser-cooling technique. Such an increase in the slow component is essential for reducing systematic and statistical errors in the precision spectroscopy of short-lifetime systems.

We quantitatively evaluated the change in the velocity distribution caused by laser cooling. We assumed a phenomenological function for the velocity profile of Ps after laser cooling and fitted it to the measured data considering the frequency resolution of the Doppler profile measurements. A fitting result is shown as the dotted curve in Fig. 2b. As conservative estimates, the best-fit value and the upper limit of the FWHM width of the decelerated component were 23 and 30 GHz, which were evaluated adopting the best resolution of 8 GHz. The corresponding temperatures were 0.8 and 1.4 K, indicating cooling of velocities equivalent to approximately 1 K or sub-kelvin temperatures. Systematic errors due to the unfixed frequency resolution varying between 8 and 16 GHz for experiment-specific reasons (see Methods) as well as statistical errors in the measurement result in a restricted precision for the evaluated Doppler broadening of the decelerated Ps. In the frequency range swept by the cooled laser, the reductions in the population of the 1S state were evaluated to be 61% and 49%, respectively. Although the unknown number of delayed Ps released from the silica aerogel precludes further quantitative discussion, most of the Ps resonating with the cooling laser was efficiently decelerated.

A control experiment was also performed, in which we set high-frequency detuning for the cooling laser. We tuned the optical frequency (detuning) to change from approximately 1,233,390 GHz (−209 GHz) to 1,233,440 GHz (−159 GHz) in 100 ns. In this case, the Ps in the probed velocity range did not resonate with the cooling laser. The open circles in Fig. 2b represent the fractional changes observed in the control experiment. The dashed line is a constant function fitted to the experimental results, indicating that there was no statistically significant change due to the largely detuned cooling laser. Thus, the cooling laser causes only a velocity change originating from the chirp cooling effect with resonant excitation and subsequent relaxation.

To gain a deeper understanding of the experimental results, we performed numerical simulations. To describe the interaction between Ps and the cooling laser and track the momentum distribution of Ps in the 1S and 2P states, we developed a framework to calculate the time evolution of the density matrix based on the Lindblad master equation (see Methods). For the time-frequency characteristics and intensity of the cooling laser pulse train, we used realistic parameters that were consistent with our experiments. In addition, we neglected interatomic interactions because of the dilute nature of the Ps gas.

Figure 3 shows how the velocity distribution of Ps in the 1S state changed over time with the cooling pulse train with a duration of 100 ns. The chirp rate of $4.9 \times 10^2$ GHz/μs corresponds to a velocity change of 120 m/s in 1 ns for Ps that resonates with the cooling laser. The single-pulse linewidth of 8.9 GHz (FWHM) corresponds to a velocity range of 2,200 m/s (FWHM) that is covered, $\Delta v_{\text{single}}$. We plotted the probability density for each velocity with respect to that at zero at the beginning of the pulse train.

The result at 42 ns (corresponding to a resonance velocity change of approximately 5,000 m/s) shows that Ps atoms with higher velocities are sequentially decelerated without velocity intermittency and accumulate on the slower side. This demonstrates the high deceleration efficiency of our cooling pulse train. At 89 ns (corresponding to a resonance velocity change of approximately 11,000 m/s), the decelerated Ps atoms were concentrated near zero velocity. The FWHM of the velocity distribution was approximately $3v_r$, and the corresponding temperature was 0.48 K.

At 127 ns, that is, 27 ns after the end of the cooling laser, most of the excited Ps atoms had relaxed to the 1S state. Considering the frequency resolution of the probe pulses and the possible presence of an uncooled fraction, the simulated distribution quantitatively reproduced the changes in the Doppler profile shown in Fig. 2b. In the velocity region where there was no interaction with the cooling laser, for example, near a velocity of 18,000 m/s, the number of Ps decayed by a factor of approximately 1/2.4 at 127 ns owing to self-decay. However, the number of Ps near zero velocity increased by a factor of three or more compared to that at 0 ns.

Note that discrete peak structures were observed in the final velocity distribution. In addition to a distribution that shows an FWHM width of approximately $v_r$ (the corresponding temperature of 55 mK) spread around the zero velocity, peaks that are shifted by $v_r$ from the zero-velocity component appeared. Although further theoretical and experimental investigations are required, we attribute this shift to the residual acceleration effects on the near-zero velocity component during the final stage of cooling. In our experiment, the central frequency of the last cooling pulse was redshifted by 9 GHz from the $1^3S_1$–$2^3P_2$ transition frequency. In this case, Ps atoms near zero velocity were accelerated because they absorb photons through $1^3S_1$–$2^3P_0$ and $1^3S_1$–$2^3P_1$ transitions. Therefore, it may be beneficial to optimise the ending frequency and intensity of the cooling laser or to perform state-selective cooling during the final stage of cooling to form a velocity distribution that shows a genuinely recoil-limited spread of a sub-100 mK temperature.

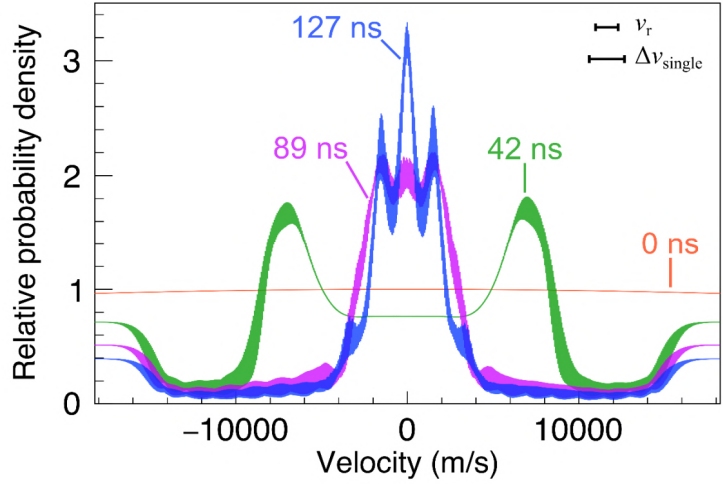

**Fig. 3: Numerical simulation of the velocity distribution of Ps based on the Lindblad master equation.** We set the duration of the cooling pulse train to 100 ns, as in the experiment. The 1D velocity distributions of 1S positronium expected at 0 ns, 42 ns, 89 ns, and 127 ns after the beginning of the cooling pulse train are shown. At 0 ns, a velocity distribution of 600 K was adopted. As the probability density of the 1S state at each velocity, we plotted relative to that of zero velocity at 0 ns. The recoil velocity $v_r$, and the velocity width $\Delta v_{\text{single}}$ that is in resonance with the FWHM frequency width of a single pulse in the cooling laser, are also shown.

We have reported the demonstration of the one-dimensional chirp cooling of Ps approaching the recoil limit. This novel chirped cooling method for Ps provides an important perspective on the use of sophisticated methodologies for conventional low-temperature atoms and molecules. The next natural extension is three-dimensional cooling, which improves the systematic error due to the second-order Doppler shift of the 1S–2S transition frequency by more than three orders of magnitude. However, there is a concern that this will result in higher attainable effective temperatures or lower decelerated populations than those reported here because the number of cooling cycles is limited by the lifetime. One solution is to use the conventional cooling method of thermal contact with low-temperature materials[29,46] as a precooling method before laser cooling.

It is necessary to increase the cooling rate by extending the cooling method to one that is not limited by spontaneous emission, as discussed for atomic and molecular cooling[47,48]. We believe that such more efficient three-dimensional cooling will enable the deceleration of most Ps atoms before annihilation, and it is important for the realisation of quantum degenerate gases of Ps, where high density and low temperature are required simultaneously. The benefits of laser cooling are the key to the realisation of BEC in this antimatter system. However, the actual required density should be fully investigated because of the lower elastic scattering rates[49] between the Ps atoms at lower temperatures, which can significantly influence the thermalisation process for inherently short-lifetime systems. When realised, comparisons with the BECs of excitons in semiconductors[50], another exotic atom system with a finite lifetime, will have important implications in quantum statistical physics.

condensates of excitons in a bulk semiconductor. *Nat. Commun.* **13**, 5388 (2022).

## Methods

**Positronium generation**

When a positron pulse is injected into a certain medium, some positrons form Ps and re-emitted as Ps into vacuum. We used these Ps in the laser-cooling experiments. Positron pulses were provided at a repetition rate of 50 Hz[43]. The positron pulse has a width of 16 ns and contains approximately $10^6$ e$^+$ per pulse. The positrons were transported with an energy of 5 keV from the positron production unit to the experimental station guided by a typical magnetic field of approximately 10 mT generated by the coils. By conducting current in the opposite direction only in the last coil immediately before the experimental station, we minimised the magnetic field in the experimental region. Iron plates for magnetic shielding and the magnetic lens located downstream contributed to this minimization, further reducing the magnetic field in the experimental region to approximately 0.15 mT. This suppresses the Zeeman effect, resulting to a negligible annihilation rate of Ps in the $1^3S_1$ state, namely ortho-Ps, due to spin mixing. The transported positrons were focused on the Ps formation medium using a magnetic lens. Approximately 1% of the incident positrons passed through the magnetic lens, and the remainder collided with the lens and were annihilated.

We used a silica aerogel, a three-dimensional network of $SiO_2$ (silica) nanograins, as the medium for the formation of Ps at room temperature. The silica aerogel had pores 45 nm in diameter, with a porosity of approximately 95%. Approximately half the positrons injected into the silica aerogel formed Ps[44]. The long-lived ortho-Ps atoms diffused and were partially released from the aerogel towards the experimental region. These ortho-Ps atoms decay to γ-rays with a lifetime of approximately 142 ns in a vacuum. γ-rays were detected using a LaBr$_3$(Ce) scintillator and a plastic scintillator, and the time-resolved γ-ray flux was measured by observing the current output of the coupled photomultiplier tubes. Note that there was no detectable influence on Ps from possible electrostatic charging of the silica aerogel.

**Laser configuration**

Ps atoms emitted into vacuum were irradiated by three different pulsed lasers. These lasers were used for chirp cooling in the 1D direction and for measuring the velocity distribution. These laser beams were incident in a direction orthogonal to the positron beam axis and reflected by a bare aluminium mirror in a counter-propagating configuration. The wavelengths of the light pulses used were 243 and 532 nm, and the reflectance of the mirror at these wavelengths was approximately 93%. The laser irradiation area was approximately 18 mm in the direction of the positron beam axis and approximately 8 mm in the vertical direction perpendicular to the positron beam axis.

The chirped pulse train laser that cooled the Ps was irradiated for approximately 100 ns after the positron pulse entered the silica aerogel. The fluence of a single pulse in an irradiated pulse train was typically 5 µJ/cm$^2$. During an irradiation duration of approximately 100 ns, the central frequency of the light pulse varied from 1,233,540 to 1,233,590 GHz. The spectral width of a single pulse was 8.9 GHz at the full width at half maximum. The cooling laser was linearly polarised with a polarization direction orthogonal that of the positron beam.

**Doppler spectroscopy**

The velocity distribution of the o-Ps in the 1S state was evaluated by Doppler spectroscopy using the 1S–2P transition. The Doppler profile of the 1S state was obtained by measuring the signal associated with the number of positrons produced by ionising Ps in the 2P state as a function of the laser frequency, which resonantly induces the 1S–2P transition. Because the Doppler shift corresponds to the velocity of Ps projected in the propagation direction of the laser beam, the velocity distribution of o-Ps can be evaluated from the measured Doppler profile. The Doppler profile measurement was performed approximately 25 ns after the end of laser cooling when the $2^3P_J$ state Ps was fully deexcited via spontaneous emission. The Doppler spectroscopy was performed at a repetition rate of 10 Hz. The laser cooling was performed at a repetition rate of 5 Hz. The change in velocity distribution associated with cooling can be evaluated by comparing the Doppler profiles with and without laser cooling.

The second harmonic of an optical parametric oscillator (OPO) excited by the third harmonic of a Q-switched Nd:YAG laser was used to induce the 1S–2P transition in the Doppler spectroscopy. The optical frequency of the second harmonic is swept around approximately 1.2336 PHz. The optical frequency was measured using a wavelength meter with an accuracy of ±3 pm (corresponding to a frequency accuracy of approximately 15 GHz). The pulse duration was approximately 3 ns. The spectral width of the second harmonic of the OPO was approximately $1.1 \times 10^2$ GHz. This spectral width was too wide to capture the changes in the velocity profile resulting from chirp cooling. Therefore, the second harmonic of the OPO was transmitted through a solid etalon to narrow the spectrum and improve the velocity resolution. The measured transmission spectral width of our custom-made solid etalon available at 243 nm varied from 8 to 16 GHz at FWHM, depending on the angle and position of incidence.

However, the Doppler broadening of Ps without cooling has an FWHM of approximately $27\sqrt{T}$ GHz at a temperature of $T$ K. This corresponds to a frequency width of 470 GHz at room temperature, which is significantly wider than the narrow resolution. To measure the Doppler profile under uncooled conditions and evaluate the temperature, it was unnecessary to spectrally narrow the second harmonic of the OPO. A larger fraction of Ps with a distributed velocity was resonant, resulting in a larger signal. Therefore, we did not employ the solid etalon when measuring the Doppler profile under uncooled conditions.

The typical incident fluence of the laser pulse that induced the 1S–2P transition was 0.27 and 2.0 μJ/cm$^2$ with and without spectral narrowing, respectively. This resulted in comparable light spectral densities for these two cases. The polarisation of the laser pulse that induced the 1S–2P transition was linear and parallel to the positron beam.

For the ionisation laser to photoionise Ps in the $2^3P_J$ state, we used the second harmonic (532 nm) of a Q-switched Nd:YAG laser. The pulse duration was 5 ns. This ionising laser pulse was delivered with the timing of the intensity peak adjusted to approximately 1.4 ns later than that of the ultraviolet (UV) nanosecond pulse, which induced the 1S–2P transition. The irradiation fluence of the 532 nm pulse was typically 15 mJ/cm$^2$. We set the ionisation laser to be linearly polarised parallel to the positron beam, similar to the laser that induced the 1S–2P transition. The repetition rate of the ionising laser was 10 Hz, which was the same as that of the laser inducing the 1S–2P transition.

Ionised positrons produced velocity selectively from the gas Ps by the two-colour pulsed lasers and were drawn into an MCP placed immediately below the interaction region between Ps and the laser. We applied a voltage of -2000 V to the input surface of the MCP and collected the ionised positrons. The MCP was sensitive to the scattered photons of deep-UV laser pulses at a wavelength of 243 nm, which resulted in a large background signal. Therefore, a pulsed negative voltage was applied to the MCP input surface. In particular, it was applied when the cooling laser irradiation was completed, with a rise time of approximately 20 ns. Therefore, the MCP gain was small at the time of incidence of deep-UV lasers, such as the cooling laser and probe laser, inducing the 1S–2P transition. This suppresses the background signal originating from photons and allows for the highly sensitive detection of ionised positrons. The voltage at the output plane of the MCP was set to 0 V. The amplified electrons were collected at a metal electrode, to which a constant voltage of +1000 V was applied. The current output from this electrode was converted to a voltage with a 50 Ω resistor, and we recorded its time evolution. The background signal originating from the deep UV photons was significantly suppressed; however, a finite contribution remained. To subtract this residual signal, the ionising laser was switched on and off every 30 s. We evaluated the signal of ionised positrons based on the difference in the integrated signals of the MCP with and without the ionising laser.

**Analysis of the measured Doppler profile of uncooled Ps**

We estimated the temperature of the Ps emitted from the silica aerogel using the measured Doppler profile. For this purpose, we defined a model function and fitted it to the data. The model function describes the number of positrons $S(\omega_R)$ generated by the photoionization process from the 2P state as a function of the central angular frequency $\omega_R$ of the probe pulse that induces the 1S–2P transition. $S(\omega_R)$ is written as

$$S(\omega_R) = \int D(v; T) \frac{A}{1 + \frac{I_S}{I(v; \omega_R)}} dv,$$

where $D(v; T)$ is the probability density of Ps with velocity $v$ and temperature $T$, where the Maxwell–Boltzmann distribution function is used; $I_S$ is the saturation intensity at the 1S–2P transition angular frequency $\omega_{eg}$, $I(v; \omega_R)$ is the light intensity at the angular frequency resonant to a Ps atom with velocity $v$. $A$ is a constant and free parameter in the fitting. The second term in the integral represents the photoionisation probability of Ps at velocity $v$. The functional form for $S(\omega_R)$ was determined using the following relation[51],

$$P_e = \frac{1}{2\left(1 + \frac{I_S}{I_R}\right)},$$

which describes the occupation probability of the excited state in a two-level system when irradiated with light of the transition frequency at an intensity $I_R$. We employed a two-level approximation because we set the spectral width of the probe pulse to be sufficiently wide compared to the splitting in the 1S–2P transition frequency. Eq. (1), determined using Eq. (2), describes the nonlinear responses to the probe laser pulse, such as the Lamb dip and saturation broadening effects in the present Doppler-broadened case.

In our experiment, we irradiated each laser beam onto Ps in a counterpropagating configuration. Therefore,

$$I(v; \omega_R) = I_L\left(\omega_{eg} + \frac{v}{c}\omega_R; \omega_R\right) + I_L\left(\omega_{eg} - \frac{v}{c}\omega_R; \omega_R\right),$$

where $I_L(\omega; \omega_R)$ is the intensity spectrum, described as a function of ω, of the probe pulse with its central angular frequency $\omega_R$. We adopted the measured spectral width of $I_L(\omega; \omega_R)$, and the intensity was a free parameter in the fitting. Here, we neglected the spatial distributions of the light intensity and Ps density. The light intensity of the probe pulse that reproduced the measurement was consistent with the actual light intensity calculated using the fluence, pulse duration, and spectral width. This result demonstrates the validity of the proposed model.

**Analysis of the fractional change in the Doppler profile**

We analysed the fractional change in the velocity distribution induced by the cooling laser by fitting the following phenomenological model to the data: The fractional change for $S_{\mathrm{on}}(f)$ and $S_{\mathrm{off}}(f)$ is defined as

$$\frac{S_{\mathrm{on}}(f) - S_{\mathrm{off}}(f)}{S_{\mathrm{off}}(f)}.$$

We first used the following raw functions that neglected the frequency resolution in the experiment:

$$S_{\mathrm{on}}^{\mathrm{raw}}(f) = \begin{cases} \exp\left(-\dfrac{m_{\mathrm{Ps}} f^2}{2 k_{\mathrm{B}} T_0 f_0^2}\right), & f < -f_{\mathrm{cooled}}, f_{\mathrm{cooled}} < f \\ A \exp\left(-\dfrac{4\log 2\, f^2}{\Delta f^2}\right) + S_{\mathrm{cooled}}, & -f_{\mathrm{cooled}} \leq f \leq f_{\mathrm{cooled}} \end{cases}$$

$$S_{\mathrm{off}}^{\mathrm{raw}}(f) = \exp\left(-\dfrac{m_{\mathrm{Ps}} f^2}{2 k_{\mathrm{B}} T_0 f_0^2}\right),$$

where the argument $f$ is the relative frequency, $f_0$ is the $1^3S_1$–$2^3P_2$ transition frequency of Ps, $m_{\mathrm{Ps}}$ is the mass of Ps, $k_{\mathrm{B}}$ is the Boltzmann constant, and $T_0$ is the temperature of Ps released from the silica aerogel. Based on the experimental results, we assumed that the Doppler profile of the uncooled Ps was a Maxwell–Boltzmann distribution at temperature $T_0 = 600$ K, based on the experimental result. The following parameters describe the change in the Doppler profile associated with cooling: These are the free parameters used in the fitting. $f_{\mathrm{cooled}}$ is the Doppler shift corresponding to the optical frequency at the beginning of the cooling laser. Furthermore, we define the Doppler width of the decelerated component as $\Delta f$. In the spectral region swept by the chirped cooling laser, the signal decreases to $S_{\mathrm{cooled}}$ by cooling. Parameter $A$ characterizes the magnitude of the decelerated component signal. These raw functions are plotted in Extended Data Fig. 1.

We generated the model functions $S_{\mathrm{on}}(f)$ and $S_{\mathrm{off}}(f)$ that correspond to the experimental results obtained by convolving $S_{\mathrm{on}}^{\mathrm{raw}}(f)$ and $S_{\mathrm{off}}^{\mathrm{raw}}(f)$ with the frequency resolution due to the linewidth of the probe pulse. The change in the Doppler profile associated with cooling was quantitatively evaluated by fitting the modelled fractional change to the measured fractional changes. In Extended Data Fig. 1, $S_{\mathrm{on}}^{\mathrm{raw}}(f)$ is plotted using the parameters obtained from the fitting.

The fitting parameters varied with the spectral width of the probe pulse, which determined the frequency resolution of the measured Doppler profile. When the spectral width, which varied in the experiment, was set to 8 GHz (narrowest), the widest Doppler spread of the cooling component was evaluated. In the main text, we have shown the corresponding best-fit value (23 GHz) and upper statistical limit (30 GHz) as conservative estimates. The upper statistical limit of the width of the cooled component was evaluated at 95% confidence level. The estimated population reductions in cooled spectral region were 61% and 49%, respectively. For a spectral width of 16 GHz for the probe pulse, the best-fit value and the upper limit of the width of the cooled component were 18 and 27 GHz, and the corresponding population reductions were 78 and 61%.

**Evaluation of the frequency resolution in the laser cooling experiment**

The frequency resolution of the fractional change in the Doppler profile as a result of the laser cooling was determined using the spectral width of the probe pulse and intensity-dependent saturation broadening. We evaluated the spectral width of the probe pulse using the optical resolution of the Fabry–Pérot solid etalon used for spectral narrowing. The FWHM optical frequency resolution as a function of the angle of incidence is shown in Extended Data Fig. 3. The resolution was evaluated by measuring the transmission spectrum of single-longitudinal-mode laser pulses at 243 nm. The spectral width of the pulses is expected to be less than 10 MHz, which is considerably narrower than the designed frequency resolution of the solid etalon; thus, enabling the evaluation of the actual resolution. We measured the transmittance as a function of the angle of incidence of the etalon. All incident angle sweeps designated in the legend were performed in the direction of increasing angle. These three sets of measurements were performed within the experimental period but not successively.

The results show that while the transmission spectral width tends to increase with the angle of incidence, it varies widely for each measurement. The degree of variation exceeds the measurement uncertainty, suggesting that the conditions of the etalon changed with each sweep experiment. The possible characteristics of the solid etalon that can cause such variations include non-uniform thickness and inhomogeneous strain on the etalon. Variations can occur because the position of the laser irradiation on the solid etalon cannot be completely fixed. To detect change in the Doppler profile resulting from laser cooling, the angles of incidence of the probe pulse on the etalon were set within the range tested above, resulting in the same degree of variation in the linewidth of the probe. Therefore, we estimated the spectral width of the probe pulse to be 8–16 GHz, based on the measured range of values shown in Extended Data Fig. 3.

Next, we examined the influence of saturation broadening, which also affects the frequency resolution. Using the effective intensity calculated from the fluence, pulse duration, and spectral width of the spectrally narrowed probe pulse, the degradation of the frequency resolution owing to saturation broadening was at most 1 GHz. Thus, saturation broadening can be neglected.

We considered the 8–16 GHz range of the frequency resolution as a systematic uncertainty in the evaluation of the fractional change. Hence, a conservative effective temperature was evaluated.

**Allowed $1^3$S–$2^3$P transitions and their intensities**

Herein, we describe the allowed transitions and their intensities among the $1^3$S–$2^3$P transitions used for laser cooling and Doppler spectroscopy. The transition matrix element is

$$-\langle n=2, L=1, S=1, J_e, M_e | \vec{d} | n=1, L=0, S=1, J_g, M_g \rangle \cdot \vec{E},$$

where $\vec{d}$ is the electric dipole moment and $\vec{E}$ is the electric field of light. $n$, $L$, and $S$ are the principal quantum number, orbital angular momentum, and total spin angular momentum, respectively. $J$ and $M$ are the total angular momentum and its projection along the quantisation axis, respectively. Subscripts e and g indicate the excited and ground states, respectively.

The electric dipole moments, when we define the quantisation axis of the atomic orbitals as the z-axis, are shown in Extended Data Fig. 2a and 2b. The direction of projection of the electric dipole moment is shown at the top of each diagram. The allowed transitions induced by the electric field of light with corresponding polarisation vectors are represented by arrows. The numbers associated with the arrows indicate the square of the absolute value of each component of the electric dipole moment normalised to the following constant:

$$|d_0|^2 = \left( \frac{128\sqrt{2}}{243} 2 e a_0 \right)^2,$$

where $e$ is the elementary charge, and $a_0$ is the Bohr radius. For some transitions, the numbers are omitted because the absolute values of the electric dipole moments coincide with those of the other transitions that differ only in the sign of $M$. The transition rates are proportional to the values shown in Extended Data Fig. 2a for linearly polarised light parallel to the z-axis, and in Extended Data Fig. 2b for orthogonally polarised light. In our experiment, the polarisations of the cooling laser pulse and probe laser pulse in Doppler spectroscopy were linear and orthogonal to each other. Extended Data Fig. 2a and 2b can be used to evaluate the transition intensity of each pulse.

Extended Data Fig. 2c shows the spontaneous emission rates from the excited states to each ground state normalised by the total decay rate $\Gamma_{\text{sp.}} \cong 3.13 \times 10^8 \text{ s}^{-1}$ from each excited state. By symmetry, the spontaneous emission rates from the states with negative $M_e$s, which are omitted from the table, are equal to the corresponding rates between the states with the signs for $M_e$ and $M_g$ reversed.

Extended Data Fig. 2a–c shows that in the cooling process, where the transitions are repeated many times, it is important to use a cooling laser with a spectral width comparable to the splitting in the transition. Otherwise, if we repeat the cooling cycle by transitioning to the $2^3P_0$, $2^3P_1$ states, for example, the $1^3S_1$ state becomes polarised and eventually making transitions to these excited states dark. In addition, the $1^3S_1$–$2^3P_2$ transition dominated the 1S–2P transitions. Therefore, we present our experimental results as functions of frequency relative to the $1^3S_1$–$2^3P_2$ frequency difference. Note that the resonance frequency observed at the one-photon transition is approximately 3 GHz higher than this frequency difference owing to the conservation laws of energy and momentum.

**Numerical simulation**

We evaluated the time evolution of the momentum distribution of Ps under the influence of a cooling laser based on the Lindblad master equation:

$$\frac{d\rho}{dt} = \frac{1}{i\hbar}[\rho, H] + L(\rho),$$

where $t$, $\hbar$, $H$ and $L(\rho)$ are the time, Dirac's constant, Hamiltonian, and Liouvillian. We considered the density matrix $\rho$ in the space spanned by the simultaneous eigenstates of the momentum of Ps and atomic configurations in the L-S coupling scheme. The interaction between Ps and the photon field was incorporated as an electric dipole interaction. This framework can describe the transitions between atomic orbitals through absorption, stimulated emission, and spontaneous emission processes, as well as momentum changes via photon recoil. We incorporated the relaxation of Ps owing to annihilation processes into the master equation as a longitudinal relaxation process.

Using the simulated velocity distribution shown in Fig. 3, we can simulate the fractional change in Fig. 2. The simulated Doppler profiles with and without irradiating the cooling laser were convolved by the spectral resolution to obtain $S_{\text{on}}^{\text{sim}}(f)$ and $S_{\text{off}}^{\text{sim}}(f)$, respectively. The spectral width of the probe pulse determines the spectral resolution. The argument *f* is the relative frequency, which is the first order Doppler shift calculated from the velocity of Ps. To express a part of the probed Ps atoms, which interacted with the cooling laser, we introduce an uncooled Ps fraction *r*. The fractional change can then be calculated as $(1-r)\frac{S_{\text{on}}^{\text{sim}}(f) - S_{\text{off}}^{\text{sim}}(f)}{S_{\text{off}}^{\text{sim}}(f)}$. The parameter *r* was determined by fitting this function to the measured data. Extended Data Fig. 4 compares the measured and simulated fractional changes. The filled circles are identical to those shown in Fig. 2b. The filled band shows the simulated result, whose interval at each relative frequency was determined by varying the frequency resolution from 8 GHz to 16 GHz. The measured data were well reproduced with the best estimated *r* ranging from 0.18 to 0.40 and the spectral resolution shown above. The statistical uncertainty of the estimated *r* is typically 0.06 in 1σ confidence level. The resultant fraction *r* is reasonable under the experimental condition, and the agreement with the measured data supports the demonstration of laser cooling of Ps.

**Data availability**

The datasets generated during and/or analysed during the current study are available from the corresponding author (yoshioka@fs.t.u-tokyo.ac.jp) on reasonable request.

**Code availability**

The codes used for modelling or analysis in the current study are available from the corresponding author (yoshioka@fs.t.u-tokyo.ac.jp) on reasonable request.

**References for Methods**

51   Foot, C. J. *Atomic Physics*.   (Oxford University Press, 2004).

**Acknowledgements**

This study was supported by the MEXT Quantum Leap Flagship Program (MEXT Q-LEAP) Grant No. JPMXS0118067246, JST FOREST Program (Grant Number JPMJFR202L), JSPS KAKENHI Grant Numbers JP16H04526, JP17H02820, JP17H06205, JP17J03691, JP18H03855, JP19H01923, MATSUO FOUNDATION, Mitutoyo Association for Science and Technology (MAST), Research Foundation for Opto-Science and Technology, and Mitsubishi Foundation. This study used the Fugaku computational resources provided by the RIKEN Centre for Computational Science through the HPCI System Research Project (Project ID: hp230215). Y. T., and R. U.

acknowledge the support of FoPM, and K. Ya also acknowledges the support of XPS, WINGS Programs, The University of Tokyo. We thank S. Uetake and K. Yoshimura (Okayama University) for providing the OPO and R. Suzuki for technical suggestions regarding the experimental setup. In addition, we thank H. Katori (The University of Tokyo), M. Kuwata-Gonokami (RIKEN), and E. Chae (Korea University) for discussions.

**Author contributions**

K. Yo. directed laser-cooling experiments. The cooling laser was conceived by K. Yo, with essential updates to the design and subsequent construction provided by K. S. The experiments were designed and performed by K. S., Y. T., R. U., N. M., and K. Yo. The laser pulses were characterised by K. S., N. M., and K. Yo. The laser system for Doppler spectroscopy in the early stages of the experiment that detected γ-rays was developed by K. Ya, and was extended with highly sensitive detection using an MCP by K. S., Y. T., and K. Yo. Data were analysed by K. S., N. M., S. S., and K. Yo. The data acquisition program was developed by K. S., A. I., S. S., and T. K. The numerical simulation was developed by K. S., R. U., S. S., and K. Yo.; A. I. led the preparation of Ps atoms. The development of γ-ray detection systems and the analysis of γ-ray data were performed by A. I., R. W. G., T. N., and S. A. Positron beam alignment was provided by K. W., I. M., and T. H. Evaluation of the silica aerogel was performed by K. S., A. I., K. I., K. M., B. E. O'R., and N. O. The magnetic lens was provided by N. O. A. I., T. N., R. W. G., N. O., K. W., and T. H. focused the positron bunch on the target aerogel and contributed to the reduction of the static magnetic field in the interaction region. A. I. and R. W. G performed a blind analysis of the laser cooling data. K. S. and K. Yo wrote the manuscript with feedback from all the authors.

**Author information**

The authors declare no competing interests. Correspondence and requests for materials should be addressed to A. I. (ishida@icepp.s.u-tokyo.ac.jp) and K. Yo. (yoshioka@fs.t.u-tokyo.ac.jp)

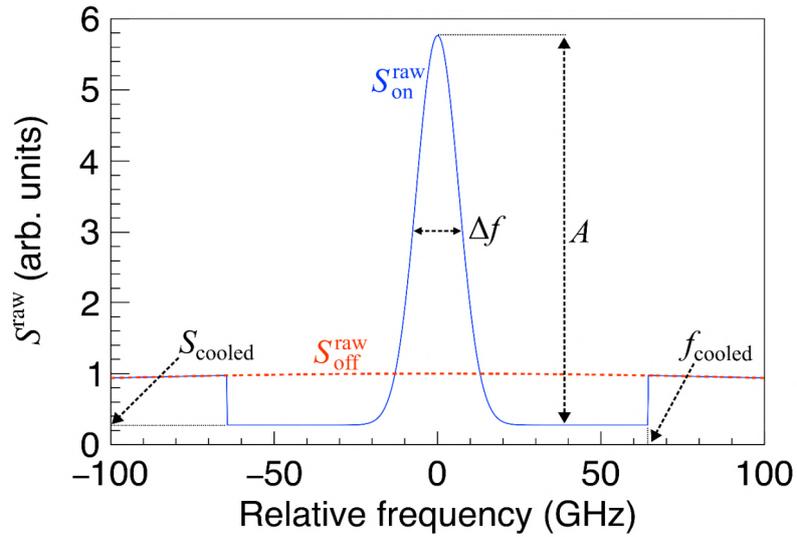

**Extended Data Fig. 1: Raw Doppler profile functions that we used to analyse the fractional change in velocity distribution associated with laser cooling.** The Doppler distribution of uncooled Ps follows the Maxwell–Boltzmann distribution, and the possible changes in the Doppler profile by cooling are characterised by parameters $A$, $S_{\text{cooled}}$, $\Delta f$, and $f_{\text{cooled}}$. Fitting of the experimental data was performed by considering the frequency resolution in Doppler spectroscopy.

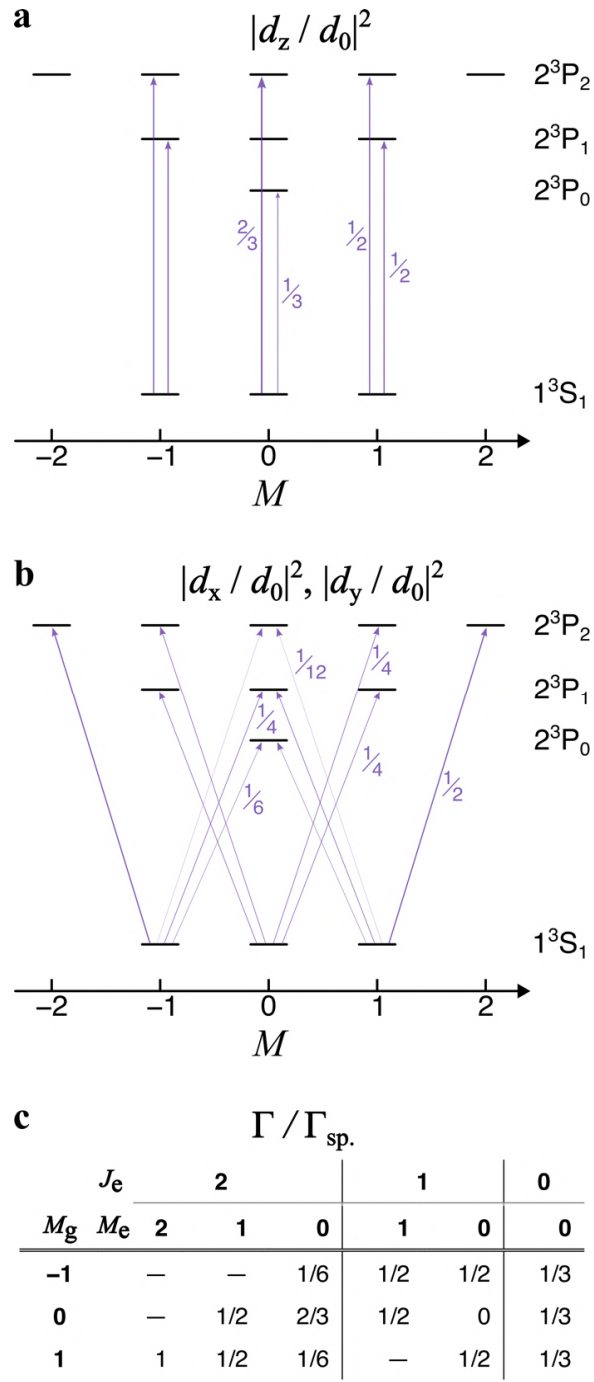

**Extended Fig. 2: Allowed $1^3S$–$2^3P$ transitions and their intensities. a, b,** Magnitudes of the electric dipole moments. **c,** Table summarising spontaneous emission rates.

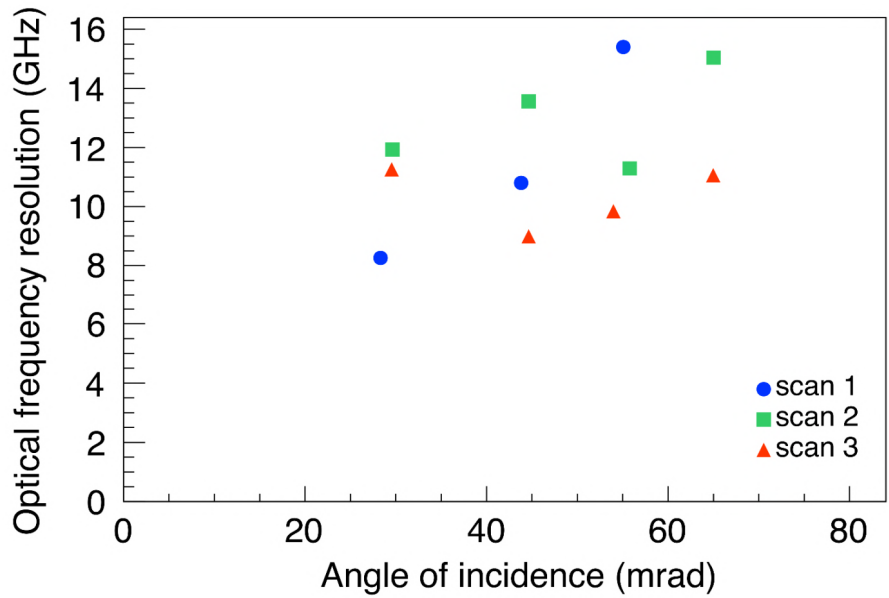

**Extended Fig. 3: Optical frequency resolution of the solid etalon used in spectral narrowing of the probe pulse for the 1S-2P transition.** We evaluated the FWHM resolution as a function of the incident angle of single-longitudinal-mode pulses. Differences in the markers represent the three scanning experiments.

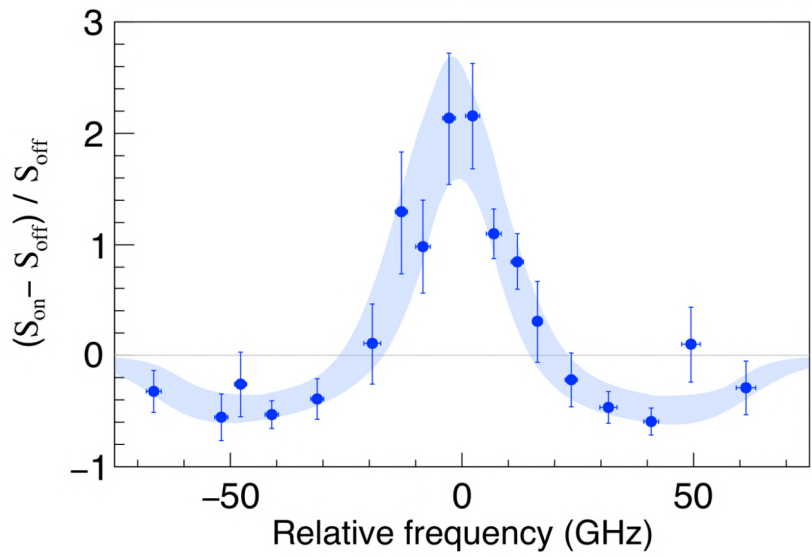

**Extended Data Fig. 4: Measured and simulated fractional change induced by the cooling laser.** The filled circles show the measured data that are identical as Fig. 2b, and the filled band shows the simulated results. The interval at each relative frequency of the band was originated from the uncertainty in the spectral width of the probe pulse.